\documentstyle[twoside,psfig]{article}

\catcode`\@=11
\long\def\@makefntext#1{
\protect\noindent \hbox to 3.2pt {\hskip-.9pt  
$^{{\eightrm\@thefnmark}}$\hfil}#1\hfill}		

\def\thefootnote{\fnsymbol{footnote}}
\def\@makefnmark{\hbox to 0pt{$^{\@thefnmark}$\hss}}	
	
\def\ps@myheadings{\let\@mkboth\@gobbletwo
\def\@oddhead{\hbox{}
\rightmark\hfil\eightrm\thepage}   
\def\@oddfoot{}\def\@evenhead{\eightrm\thepage\hfil
\leftmark\hbox{}}\def\@evenfoot{}
\def\sectionmark##1{}\def\subsectionmark##1{}}

\oddsidemargin=\evensidemargin
\renewcommand{\thefootnote}{\fnsymbol{footnote}}

\newcounter{sectionc}\newcounter{subsectionc}\newcounter{subsubsectionc}
\renewcommand{\section}[1] {\vspace{12pt}\addtocounter{sectionc}{1} 
\setcounter{subsectionc}{0}\setcounter{subsubsectionc}{0}\noindent 
	{\tenbf\thesectionc. #1}\par\vspace{5pt}}
\renewcommand{\subsection}[1] {\vspace{12pt}\addtocounter{subsectionc}{1} 
	\setcounter{subsubsectionc}{0}\noindent 
	{\bf\thesectionc.\thesubsectionc. {\kern1pt \bfit #1}}\par\vspace{5pt}}
\renewcommand{\subsubsection}[1] {\vspace{12pt}\addtocounter{subsubsectionc}{1}
	\noindent{\tenrm\thesectionc.\thesubsectionc.\thesubsubsectionc.
	{\kern1pt \tenit #1}}\par\vspace{5pt}}
\newcommand{\nonumsection}[1] {\vspace{12pt}\noindent{\tenbf #1}
	\par\vspace{5pt}}

\newcounter{appendixc}
\newcounter{subappendixc}[appendixc]
\newcounter{subsubappendixc}[subappendixc]
\renewcommand{\thesubappendixc}{\Alph{appendixc}.\arabic{subappendixc}}
\renewcommand{\thesubsubappendixc}
	{\Alph{appendixc}.\arabic{subappendixc}.\arabic{subsubappendixc}}

\renewcommand{\appendix}[1] {\vspace{12pt}
        \refstepcounter{appendixc}
        \setcounter{figure}{0}
        \setcounter{table}{0}
        \setcounter{lemma}{0}
        \setcounter{theorem}{0}
        \setcounter{corollary}{0}
        \setcounter{definition}{0}
        \setcounter{equation}{0}
        \renewcommand{\thefigure}{\Alph{appendixc}.\arabic{figure}}
        \renewcommand{\thetable}{\Alph{appendixc}.\arabic{table}}
        \renewcommand{\theappendixc}{\Alph{appendixc}}
        \renewcommand{\thelemma}{\Alph{appendixc}.\arabic{lemma}}
        \renewcommand{\thetheorem}{\Alph{appendixc}.\arabic{theorem}}
        \renewcommand{\thedefinition}{\Alph{appendixc}.\arabic{definition}}
        \renewcommand{\thecorollary}{\Alph{appendixc}.\arabic{corollary}}
        \renewcommand{\theequation}{\Alph{appendixc}.\arabic{equation}}
        \noindent{\tenbf Appendix \theappendixc #1}\par\vspace{5pt}}
\newcommand{\subappendix}[1] {\vspace{12pt}
        \refstepcounter{subappendixc}
        \noindent{\bf Appendix \thesubappendixc. {\kern1pt \bfit #1}}
	\par\vspace{5pt}}
\newcommand{\subsubappendix}[1] {\vspace{12pt}
        \refstepcounter{subsubappendixc}
        \noindent{\rm Appendix \thesubsubappendixc. {\kern1pt \tenit #1}}
	\par\vspace{5pt}}

\newcommand{\textlineskip}{\baselineskip=13pt}
\newcommand{\smalllineskip}{\baselineskip=10pt}

\def\eightcirc{
\begin{picture}(0,0)
\put(4.4,1.8){\circle{6.5}}
\end{picture}}
\def\eightcopyright{\eightcirc\kern2.7pt\hbox{\eightrm c}} 

\newcommand{\copyrightheading}[1]
	{\vspace*{-2.5cm}\smalllineskip{\flushleft
	{\footnotesize Modern Physics Letters A, #1}\\
	{\footnotesize $\eightcopyright$\, World Scientific Publishing
	 Company}\\
	 }}

\def\abstracts#1#2#3{{
	\centering{\begin{minipage}{4.5in}\footnotesize\baselineskip=10pt
	\parindent=0pt #1\par 
	\parindent=15pt #2\par
	\parindent=15pt #3
	\end{minipage}}\par}} 

\def\keywords#1{{
	\centering{\begin{minipage}{4.5in}\footnotesize\baselineskip=10pt
	{\footnotesize\it Keywords}\/: #1
	 \end{minipage}}\par}}

\newcommand{\bibit}{\nineit}
\newcommand{\bibbf}{\ninebf}
\renewenvironment{thebibliography}[1]
	{\frenchspacing
	 \ninerm\baselineskip=11pt
	 \begin{list}{\arabic{enumi}.}
        {\usecounter{enumi}\setlength{\parsep}{0pt}     
	 \setlength{\leftmargin 12.7pt}{\rightmargin 0pt} 
         \setlength{\itemsep}{0pt} \settowidth
	{\labelwidth}{#1.}\sloppy}}{\end{list}}

\newcounter{itemlistc}
\newcounter{romanlistc}
\newcounter{alphlistc}
\newcounter{arabiclistc}
\newenvironment{itemlist}
    	{\setcounter{itemlistc}{0}
	 \begin{list}{$\bullet$}
	{\usecounter{itemlistc}
	 \setlength{\parsep}{0pt}
	 \setlength{\itemsep}{0pt}}}{\end{list}}

\newenvironment{romanlist}
	{\setcounter{romanlistc}{0}
	 \begin{list}{$($\roman{romanlistc}$)$}
	{\usecounter{romanlistc}
	 \setlength{\parsep}{0pt}
	 \setlength{\itemsep}{0pt}}}{\end{list}}

\newcommand{\fcaption}[1]{
        \refstepcounter{figure}
        \setbox\@tempboxa = \hbox{\footnotesize Fig.~\thefigure. #1}
        \ifdim \wd\@tempboxa > 5in
           {\begin{center}
        \parbox{5in}{\footnotesize\smalllineskip Fig.~\thefigure. #1}
            \end{center}}
        \else
             {\begin{center}
             {\footnotesize Fig.~\thefigure. #1}
              \end{center}}
        \fi}

\newcommand{\tcaption}[1]{
        \refstepcounter{table}
        \setbox\@tempboxa = \hbox{\footnotesize Table~\thetable. #1}
        \ifdim \wd\@tempboxa > 5in
           {\begin{center}
        \parbox{5in}{\footnotesize\smalllineskip Table~\thetable. #1}
            \end{center}}
        \else
             {\begin{center}
             {\footnotesize Table~\thetable. #1}
              \end{center}}
        \fi}

\def\@citex[#1]#2{\if@filesw\immediate\write\@auxout
	{\string\citation{#2}}\fi
\def\@citea{}\@cite{\@for\@citeb:=#2\do
	{\@citea\def\@citea{,}\@ifundefined
	{b@\@citeb}{{\bf ?}\@warning
	{Citation `\@citeb' on page \thepage \space undefined}}
	{\csname b@\@citeb\endcsname}}}{#1}}

\newif\if@cghi
\def\cite{\@cghitrue\@ifnextchar [{\@tempswatrue
	\@citex}{\@tempswafalse\@citex[]}}
\def\citelow{\@cghifalse\@ifnextchar [{\@tempswatrue
	\@citex}{\@tempswafalse\@citex[]}}
\def\@cite#1#2{{$\null^{#1}$\if@tempswa\typeout
	{IJCGA warning: optional citation argument 
	ignored: `#2'} \fi}}

\def\pmb#1{\setbox0=\hbox{#1}
	\kern-.025em\copy0\kern-\wd0
	\kern.05em\copy0\kern-\wd0
	\kern-.025em\raise.0433em\box0}

\def\fnt#1#2{\footnotetext{\kern-.3em
	{$^{\mbox{\scriptsize #1}}$}{#2}}}

\def\fpage#1{\begingroup
\voffset=.3in
\thispagestyle{empty}\begin{table}[b]\centerline{\footnotesize #1}
	\end{table}\endgroup}

\font\tenrm=cmr10
\font\tenit=cmti10 
\font\tenbf=cmbx10
\font\bfit=cmbxti10 at 10pt
\font\ninerm=cmr9
\font\nineit=cmti9
\font\ninebf=cmbx9
\font\eightrm=cmr8

\def\qed{\hbox{${\vcenter{\vbox{			
   \hrule height 0.4pt\hbox{\vrule width 0.4pt height 6pt
   \kern5pt\vrule width 0.4pt}\hrule height 0.4pt}}}$}}

\renewcommand{\thefootnote}{\fnsymbol{footnote}}	

\begin{document}
\newcommand{\pyth}{\textsc{Py\-thia}}
\newcommand{\pythjet}{\textsc{Py\-thia/Jet\-set}}
\newcommand{\cerenkov}{\v{C}erenkov}
\newcommand{\eg}{e.g.}
\setlength{\textheight}{7.7truein}  

\normalsize\textlineskip
\thispagestyle{empty}
\setcounter{page}{1}

\copyrightheading{}			

\vspace*{0.46truein}
\begin{flushright}
    \small FERMILAB-Pub-00/200-E \\ UMS/HEP/2000--025~~~~~~~~~
\end{flushright}
\vspace*{0.10truein}

\fpage{1}
\centerline{\bf Review of Recent Searches for Rare and Forbidden}
\baselineskip=13pt
\centerline{\bf Dilepton Decays of Charmed Mesons}
\vspace*{0.37truein}
\centerline{\footnotesize David A. Sanders}
\baselineskip=12pt
\centerline{\footnotesize\it Department of Physics and Astronomy, 
University of Mississippi--Oxford, 122 Lewis Hall}
\baselineskip=10pt
\centerline{\footnotesize\it University, MS 38677, USA}
\vspace*{22pt}


\vspace*{0.21truein}
\abstracts{I briefly review the results of recent searches for 
flavor-changing neutral current and lepton-flavor and lepton-number 
violating decays of $D^+$, $D_{s}^{+}$, and $D^0$ mesons (and their 
antiparticles) into modes containing muons and electrons. The primary focus 
is the results from Fermilab charm hadroproduction experiment E791. 
E791 examined 24 $\pi \ell \ell$ and $K\ell \ell$ decay modes of $D^+$ 
and $D_{s}^{+}$ and $\ell^+ \ell^-$ decay modes of $D^0$.  Limits presented 
by E791 for 22 rare and forbidden dilepton decays of $D$ mesons were 
more stringent than those obtained from previous searches,\cite{PDG} or 
else were the first reported.}{}{}

\vspace*{10pt}
\keywords{Charm, Rare, Forbidden, Decay, Dilepton}


\vspace*{1pt}\textlineskip	
\section{Introduction}	
\vspace*{-0.5pt}
\noindent
Searches for rare and forbidden decays allow investigation of physics 
beyond the Standard Model and of phenomena in mass ranges beyond those 
available to current accelerators.  One way to search for physics beyond 
the Standard Model is to examine decay modes that are forbidden or else are 
predicted to occur at a negligible level (rare). Observing such decays would 
constitute evidence for new physics, and measuring their branching fractions 
would provide insight into how to modify our theoretical understanding, \eg, 
by introducing new particles or new gauge couplings. These new mediators 
could include: leptoquarks, horizontal gauge bosons, or something 
even more esoteric.

This review focuses on recent results from the Fermilab experiment  
E791\cite{FCNCNew} search for 24 decay modes of 
the neutral and charged $D$ mesons (which contain the heavy charm 
quark). These decay modes\footnote{Charge-conjugate modes are 
included implicitly throughout this paper.} ~fall into three categories:
\begin{romanlist}
\item FCNC -- flavor-changing neutral current decays 
($D^0\rightarrow\ell^+\ell^-$ and
$D^+_{(d,s)}\rightarrow h^+\ell^+\ell^-$, in which $h$ is either a 
$\pi$ or $K$);
\item LFV -- lepton-flavor violating decays 
($D^{0}\rightarrow \mu^{\pm }e^{\mp }$,
$D^+_{(d,s)}\rightarrow h^+\mu^{\pm }e^{\mp }$, and 
$D^+_{(d,s)}\rightarrow h^-\mu^+e^+$,
in which the leptons belong to different generations);
\item LNV -- lepton-number violating decays
($D^+_{(d,s)}\rightarrow h^-\ell^+\ell^+$,
in which the leptons belong to the same generation
but have the same sign charge).
\end{romanlist}
Decay modes belonging to (i) occur within the Standard Model
via higher-order diagrams, but estimated branching fractions\cite{SCHWARTZ93} 
are $10^{-8}$ to $10^{-6}$. Such small rates are 
below the sensitivity of current experiments. However, if 
additional particles such as supersymmetric squarks or charginos exist, 
they could contribute additional amplitudes that would make these modes 
observable. Decay modes belonging to (ii) and (iii) do not conserve lepton 
number and thus are forbidden within the Standard Model. However, lepton 
number conservation is not required by Lorentz invariance or gauge 
invariance, and a number of theoretical extensions to the Standard 
Model predict lepton-number violation.\cite{Pakvasa} Many experiments 
have searched for lepton-number violation in $K$ decays,\cite{Kref} and for 
lepton-number violation and flavor-changing neutral currents in 
$D$ \cite{FCNC}$^{-}$\cite{CLEO} and $B$ \cite{Bref} decays. While many 
experiments have examined charge 1/3 strange and beauty quarks, 
investigations of $D$ mesons look for rare and forbidden decays involving 
charge 2/3 charm quarks. Charge 2/3 quarks may couple differently than 
charge 1/3 quarks.\cite{Pakvasa}
\setcounter{footnote}{0}
\renewcommand{\thefootnote}{\alph{footnote}}

\section{E791 Analysis}
\noindent
The data were gathered by the the E791 spectrometer,\cite{e791spect} which 
recorded $2 \times 10^{10}$ events with a loose transverse energy 
trigger. These events were produced by a 500 GeV/$c$~ $\pi ^{-}$ beam 
interacting in a fixed target consisting of five thin, well separated foils.
Track and vertex reconstruction was provided by 23 silicon microstrip 
planes\cite{SMD} and 45 wire chamber planes plus, two magnets.

Electron identification (ID) was based on transverse shower shape plus 
matching wire chamber tracks to shower positions and energies in the 
electromagnetic calorimeter.\cite{SLIC} The electron ID efficiency 
varied from 62$\%$ below 9 GeV to 45$\%$ above 20 GeV. The probability 
to mis-ID a pion as an electron was about 0.8$\%$.

Muon identification was obtained from two planes of scintillation 
counters. The first plane (5.5 m $\times$ 3.0 m) of 15 counters 
measured the horizontal $x$ axis while the second plane (3.0 m $\times$ 
2.2 m) of 16 counters measured the vertical $y$ axis. The counters had 
about 15 interaction lengths of shielding.
Candidate muon tracks were required to pass cuts that were set using
$D^+\rightarrow \overline{K}^{*0} \mu^{+}\nu _{\!_{\mu}}$ decays from 
E791's data.\cite{Chong} Timing from the $y$ counters improved  
$x$ position resolution. Counter efficiencies were measured using muons 
originating from the primary beam dump, and were found to be $(99\pm 1)\%$ 
for the $y$ counters and $(69\pm 3)\%$ for the $x$ counters. The probability 
for misidentifying a pion as a muon decreased with increasing momentum; 
from about 6$\%$ at 8 GeV/$c$ to $(1.3 \pm 0.1)\%$ above 20 GeV/$c$.

After reconstruction,\cite{FARM791} events with evidence of well-separated 
production (primary) and decay (secondary) vertices were selected to 
separate charm candidates from background.  Secondary vertices had to 
be separated from the primary vertex by greater than 
$20\,\sigma_{_{\!L}}$ for $D^+$ decays and greater than 
$12\,\sigma_{_{\!L}}$ for $D^0$ and $D_{s}^{+}$ decays, 
where $\sigma_{_{\!L}}$ was the calculated resolution of the measured 
longitudinal separation. In addition, the secondary vertex had to be 
separated from the closest material in the target foils by greater than 
$5\,\sigma_{_{\!L}}^{\prime }$, where $\sigma_{_{\!L}}^{\prime }$ was 
the uncertainty in this separation. The sum of the vector momenta of 
the tracks from the secondary vertex was required to pass within 
$40~\mu$m of the primary vertex in the plane perpendicular to the beam. 
Finally, the net momentum of the charm candidate transverse to 
the line connecting the production and decay vertices had to be 
less than 300, 250, and 200 MeV/$c$ for $D^0$, $D_{s}^{+}$, and $D^+$
candidates, respectively. The decay track candidates passed 
approximately 10 times closer to the secondary vertex than to the 
primary vertex. These selection criteria and, where possible, 
the kaon identification\cite{Bartlett} requirements, were the same for 
the search mode and for its normalization signal.

For this study a ``blind'' analysis technique was used. Before the 
selection criteria were finalized, all events having masses within a 
mass window $\Delta M_S$ around the mass of $D^{+}$, 
$D_{s}^{+}$, or $D^{0}$ were ``masked'' so that the presence or 
absence of any potential signal candidates would not bias the choice of 
selection criteria. All criteria were then chosen by studying 
signal events generated by a Monte Carlo simulation program (see 
below) and background events, outside the signal windows, from real data. 
Events within the signal windows were unmasked only after this 
optimization. Background events were chosen from a mass window 
$\Delta M_B$ above and below the signal window $\Delta M_S$. The criteria 
were chosen to maximize the ratio $N_S/\sqrt{N_B}$, where $N_S$ and $N_B$ 
are the numbers of signal and background events, respectively. Asymmetric 
windows for the decay modes containing electrons were used to allow for 
the bremsstrahlung low-energy tail. The signal windows were: 

\begin{itemlist}
 \item $1.84<M(D^{+})<1.90~{\rm GeV}/c^{\,2}~{\rm for}~ 
 D^{+}\rightarrow h\mu \mu $,
 \item $1.78<M(D^{+})<1.90 ~{\rm GeV}/c^{\,2}~{\rm for}~ 
 D^{+}\rightarrow hee ~{\rm and}~  h\mu e$,
 \item $1.95<M(D_{s}^{+})<1.99 ~{\rm GeV}/c^{\,2}~{\rm for}~ 
 D_{s}^{+}\rightarrow h\mu \mu $,
 \item $1.91<M(D_{s}^{+})<1.99 ~{\rm GeV}/c^{\,2}~{\rm for}~ 
 D_{s}^{+}\rightarrow hee~{\rm and}~ h\mu e$,
 \item $1.83<M(D^{0})<1.90 ~{\rm GeV}/c^{\,2}~{\rm for}~ 
 D^{0}\rightarrow \mu \mu $,
 \item $1.76<M(D^{0})<1.90 ~{\rm GeV}/c^{\,2}~{\rm for}~ 
 D^{0}\rightarrow ee~{\rm and}~ \mu e$.
\end{itemlist}

The sensitivity of the search was normalized to topologically similar 
Cabibbo-favored decays. For $D^{+}$, $24010\pm 166$ 
$D^+\rightarrow K^-\pi^+\pi^+$ decays were used; for $D_{s}^{+}$, 
$782\pm 30$ $D_{s}^{+}\rightarrow \phi \pi^+$ decays were used; and for 
$D^{0}$, $25210\pm 179$ $D^0\rightarrow K^-\pi^+$ decays were used. The 
normalization modes widths were 10.5 MeV/$c^{\,2}$ for $D^{+}$, 9.5 
MeV/$c^{\,2}$ for $D_{s}^{+}$, and 12 MeV/$c^{\,2}$ for $D^{0}$.  
The upper limit for each branching fraction $B_{X}$ was calculated 
using the following formula:
$B_{X} =  
({N_{X}}/{N_{\mathrm{Norm}}}) \cdot
({\varepsilon _{\mathrm{Norm}}}/{\varepsilon _{X}})
\cdot B_{\mathrm{Norm}}$,
where $N_{X}$ was the 90$\%$ CL upper limit on the number of decays 
for rare or forbidden decay mode $X$, and $\varepsilon_{X}$ was that 
mode's detection efficiency. $N_{\mathrm{Norm}}$ was the fitted number 
of normalization mode decays; $\varepsilon_{\mathrm{Norm}}$ 
was the normalization mode detection efficiency; and 
$B_{\mathrm{Norm}}$ was the normalization mode branching fraction 
obtained from the Particle Data Group.\cite{PDG} 

The ratio of detection efficiencies is given by 
${\varepsilon _{\mathrm{Norm}}}/{\varepsilon _{X}} =
{N_{\mathrm{Norm}}^{\mathrm{MC}}}/{N_{X}^{\mathrm{MC}}}$,
where $N_{\mathrm{Norm}}^{\mathrm{MC}}$ and $N_{X}^{\mathrm{MC}}$ are 
the fractions of Monte Carlo events that are reconstructed and pass 
final cuts, for the normalization and decay modes, 
respectively. We use \pythjet\cite{MC} as the physics 
generator and model effects of resolution, geometry, magnetic 
fields, multiple scattering, interactions in detector material, 
detector efficiencies, and analysis cuts. The 
efficiencies for the normalization modes varied from about 
$0.5\%$ to $2\%$ and the 
search modes varied from about $0.1\%$ to $2\%$. 

Monte Carlo studies showed that the experiment's acceptances are nearly 
uniform across the Dalitz plots, except that the dilepton 
identification efficiencies typically dropped to near zero at the 
dilepton mass threshold. While the loss in efficiency varied channel by 
channel, the efficiency typically reached its full value at masses 
only a few hundred MeV/$c^{\,2}$ above the dilepton mass threshold. 
A constant weak-decay matrix element was used when calculating the overall 
detection efficiencies. Two exceptions to the use of the Monte Carlo 
simulations in determining relative efficiencies were made: those for 
\cerenkov{} identification when the number of kaons in the signal and 
normalization modes were different, and those for the muon 
identification. These efficiencies were determined from data.

\section{E791 Results}
\noindent
The 90$\%$ CL upper limits $N_{X}$ were calculated using the method of 
Feldman and Cousins\cite{Cousins} to account for background, and then 
corrected for systematic errors by the method of Cousins and 
Highland.\cite{COUSINSHI} In these methods, the numbers of signal events 
are determined by simple counting, not by a fit. All results are listed 
in Table 1 and shown in Fig. 1. The kinematic criteria and removal of 
reflections (see below) were different for the $D^{+}$, $D_{s}^{+}$, 
and $D^0$. 

The upper limits were determined by both the number of candidate events 
and the expected number of background events within the signal region.  
Background sources that were not removed by the selection 
criteria discussed earlier included decays in which hadrons (from 
real, fully-hadronic decay vertices) were misidentified as leptons. 
In the case where kaons were misidentified as leptons, candidates 
had effective masses which lie outside the signal windows. Most of 
these originated from the Cabibbo-favored modes 
$D^+\rightarrow K^-\pi^+\pi^+$, $D_{s}^{+}\rightarrow K^-K^+\pi^+$, 
and $D^0\rightarrow K^-\pi^+$ (and charge conjugates). These 
Cabibbo-favored reflections were explicitly removed prior to the 
selection-criteria optimization. There remained two sources of background 
in the data: hadronic decays with pions misidentified as leptons 
($N_{\mathrm{MisID}}$) and ``combinatoric'' background 
($N_{\mathrm{Cmb}}$) arising primarily from false vertices and 
partially reconstructed charm decays. After selection criteria 
were applied and the signal windows opened, the number of events 
within the window was $N_{\mathrm{Obs}} = N_{\mathrm{Sig}}+ 
N_{\mathrm{MisID}} + N_{\mathrm{Cmb}}$. 

The background $N_{\mathrm{MisID}}$ arose mainly 
from singly-Cabibbo-suppressed (SCS) modes. These misidentified 
leptons can come from hadronic showers 
reaching the muon counter, decays-in-flight, and random overlaps of 
tracks from otherwise separate decays (``accidental'' sources). A limit 
for $D^+\rightarrow K^-\ell^+\ell^+$ modes was not attempted, as they had 
relatively large feedthrough signals from copious 
Cabibbo-favored $K^-\pi^+\pi^+$ decays. Instead, the observed signals in 
$K^-\ell^+\ell^+$ channels were used to measure three dilepton 
misidentification rates under the assumption that the observed signals 
arise entirely from lepton misidentification.
\begin{figure}[h!] 
\vspace*{-56pt}
\centerline{\psfig{file=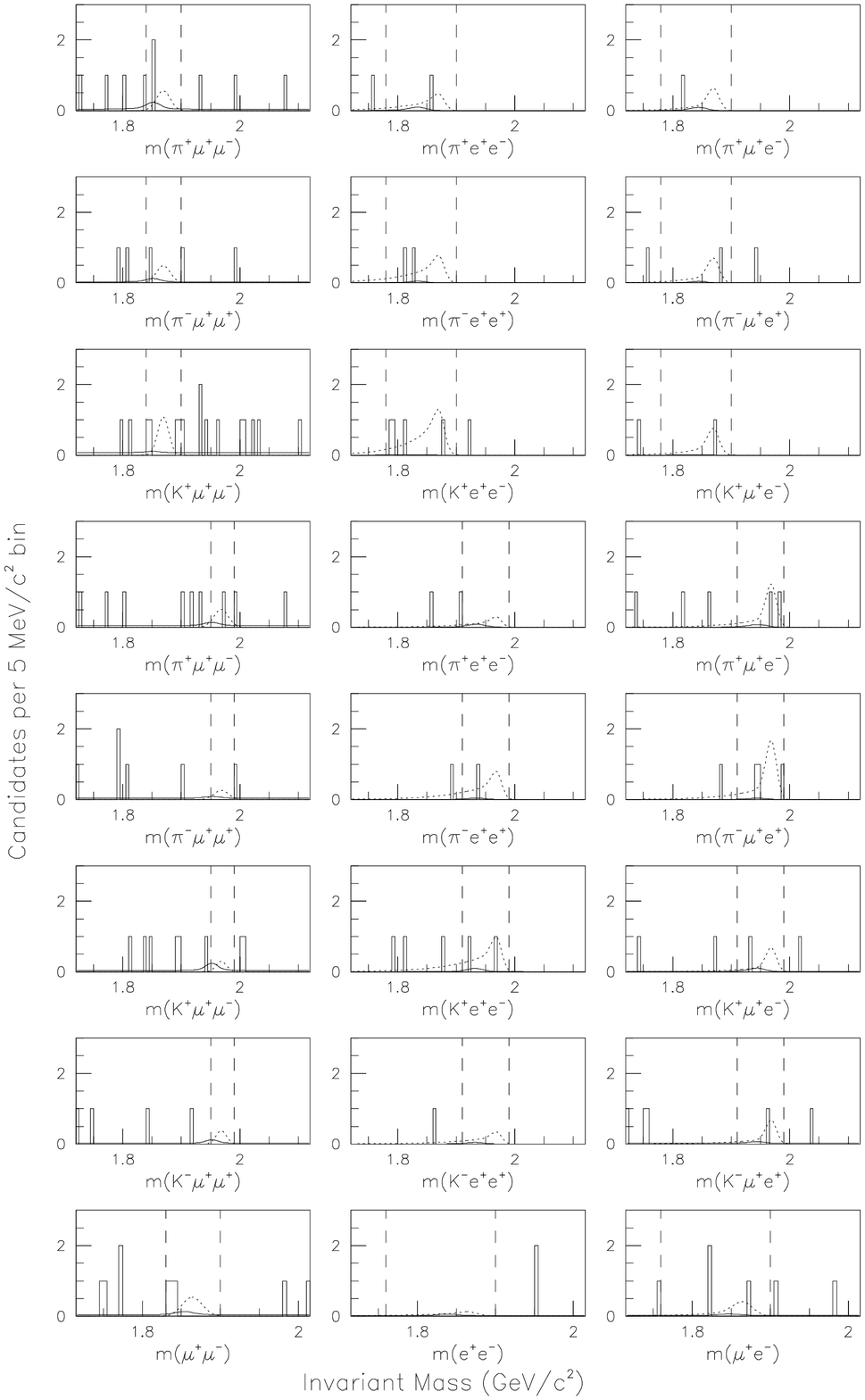,width=4truein}} 
\vspace*{3pt}
\fcaption{
\small Final event samples for $D^+$ (rows 1--3), 
$D_{s}^{+}$ (rows 4--7), and $D^0$ (row 8) decays. Solid curves 
represent estimated background; dotted curves represent signal 
shape of a number of events equal to a 90$\%$ CL upper limit. 
Dashed vertical lines are $\Delta M_S$ windows.}
\end{figure}

The following misidentification rates were obtained: 
$r_{\mu\mu}= (7.3 \pm 2.0)\times 10^{-4}$, 
$r_{\mu e}= (2.9 \pm 1.3 )\times 10^{-4}$, and 
$r_{e e}= (3.4 \pm 1.4)\times 10^{-4}$. 
Using these rates the numbers of misidentified candidates, 
$N_{\mathrm{MisID}}^{h\ell\ell}$ (for $D^{+}$ and $D_{s}^{+}$) and 
$N_{\mathrm{MisID}}^{\ell\ell}$ (for $D^{0}$), in the signal windows 
were estimated as follows:
$N_{\mathrm{MisID}}^{h\ell\ell} = r_{\ell\ell} 
\cdot N_{\mathrm{SCS}}^{h\pi\pi}$
and
$N_{\mathrm{MisID}}^{\ell\ell} = r_{\ell\ell} 
\cdot N_{\mathrm{SCS}}^{\pi\pi}$,
where $N_{\mathrm{SCS}}^{h\pi\pi}$ and $N_{\mathrm{SCS}}^{\pi\pi}$ 
were the numbers of SCS hadronic decay candidates within the signal 
windows. For modes in which two possible pion combinations can 
contribute, \eg, $D^+\rightarrow \pi ^{+}\mu ^{\pm}\mu ^{\mp}$, 
twice the above rate was used.

To estimate the combinatoric background $N_{\mathrm{Cmb}}$ within a 
signal window $\Delta M_S$, events having masses within an 
adjacent background mass window $\Delta M_B$ were counted, and this number 
($N_{\Delta M_B}$) was scaled by the relative sizes of these windows:
$N_{\mathrm{Cmb}} = ({\Delta M_S}/{\Delta M_B}) \cdot N_{\Delta M_B}$. 
To be conservative in calculating the 90$\%$ confidence level upper 
limits, combinatoric backgrounds were taken to be zero when no 
events are located above the mass windows. In Table 1 the numbers of 
combinatoric background, misidentification background, and observed 
events are presented for all 24 modes. Also previously 
published results are given for comparison. 

\begin{table}[h!]
\tcaption{
E791 90$\%$ confidence level (CL) branching fractions (BF) compared 
to previous limits. The background and candidate events 
correspond to the signal region only.}
\vskip 5pt
\tabcolsep=2.93pt
\begin{tabular}{lcccccllr} \hline
&\small{(Est.}&\small{BG)}&\small{Cand.}&\small{Sys.}&\small{90$\%$ CL}&
\small{E791}&\small{Previous}&\\
Mode&\small{$N_{\mathrm{Cmb}}$}&\small{$N_{\mathrm{MisID}}$}&\small{Obs.}&
\small{Err.}&\small{Num.}&\small{$BF$ Limit}&\small{$BF$ Limit}&
\small{Ref.}\\
\hline
\vspace*{-11pt} &     &   &     &     &       &    &  & \\
 $D^{+}\!\rightarrow \pi ^{+}\mu ^{+}\mu ^{-}$&1.20&1.47&2&10$\%$&3.35
 &$1.5\times 10^{-5}$&$1.8\times 10^{-5}$&\cite{FCNC}\\
 $D^{+}\!\rightarrow \pi ^{+}e^{+}e^{-}$&0.00&0.90&1&12$\%$&3.53
 &$5.2\times 10^{-5}$&$6.6\times 10^{-5}$&\cite{FCNC}\\
 $D^{+}\!\rightarrow \pi ^{+}\mu ^{\pm }e^{\mp }$&0.00&0.78&1&11$\%$&3.64
 &$3.4\times 10^{-5}$&$1.2\times 10^{-4}$&\cite{E687}\\
 $D^{+}\!\rightarrow \pi ^{-}\mu ^{+}\mu ^{+}$&0.80&0.73&1&9$\%$&2.92
 &$1.7\times 10^{-5}$&$8.7\times 10^{-5}$&\cite{E687}\\
 $D^{+}\!\rightarrow \pi ^{-}e^{+}e^{+}$&0.00&0.45&2&12$\%$&5.60
 &$9.6\times 10^{-5}$&$1.1\times 10^{-4}$&\cite{E687}\\
 $D^{+}\!\rightarrow \pi ^{-}\mu ^{+}e^{+}$&0.00&0.39&1&11$\%$&4.05
 &$5.0\times 10^{-5}$&$1.1\times 10^{-4}$&\cite{E687}\\
 $D^{+}\!\rightarrow K^{+}\mu ^{+}\mu ^{-}$&2.20&0.20&3&8$\%$&5.07
 &$4.4\times 10^{-5}$&$9.7\times 10^{-5}$&\cite{E687}\\
 $D^{+}\!\rightarrow K^{+}e^{+}e^{-}$&0.00&0.09&4&11$\%$&8.72
 &$2.0\times 10^{-4}$&$2.0\times 10^{-4}$&\cite{E687}\\
 $D^{+}\!\rightarrow K^{+}\mu ^{\pm }e^{\mp }$&0.00&0.08&1&9$\%$&4.34
 &$6.8\times 10^{-5}$&$1.3\times 10^{-4}$&\cite{E687}\\
\hline
\vspace*{-11pt} &     &   &     &     &         &    & & \\
 $D_{s}^{+}\!\rightarrow K^{+}\mu ^{+}\mu ^{-}$&0.67&1.33&0&27$\%$&1.32
 &$1.4\times 10^{-4}$&$5.9\times 10^{-4}$&\cite{E653}\\
 $D_{s}^{+}\!\rightarrow K^{+}e^{+}e^{-}$&0.00&0.85&2&29$\%$&5.77
 &$1.6\times 10^{-3}$&\\
 $D_{s}^{+}\!\rightarrow K^{+}\mu ^{\pm }e^{\mp }$&0.40&0.70&1&27$\%$&3.57
 &$6.3\times 10^{-4}$&&\\
 $D_{s}^{+}\!\rightarrow K^{-}\mu ^{+}\mu ^{+}$&0.40&0.64&0&26$\%$&1.68
 &$1.8\times 10^{-4}$&$5.9\times 10^{-4}$&\cite{E653}\\
 $D_{s}^{+}\!\rightarrow K^{-}e^{+}e^{+}$&0.00&0.39&0&28$\%$&2.22
 &$6.3\times 10^{-4}$&&\\
 $D_{s}^{+}\!\rightarrow K^{-}\mu ^{+}e^{+}$&0.80&0.35&1&27$\%$&3.53
 &$6.8\times 10^{-4}$&&\\
 $D_{s}^{+}\!\rightarrow \pi ^{+}\mu ^{+}\mu ^{-}$&0.93&0.72&1&27$\%$&3.02
 &$1.4\times 10^{-4}$&$4.3\times 10^{-4}$&\cite{E653}\\
 $D_{s}^{+}\!\rightarrow \pi ^{+}e^{+}e^{-}$&0.00&0.83&0&29$\%$&1.85
 &$2.7\times 10^{-4}$&&\\
 $D_{s}^{+}\!\rightarrow \pi ^{+}\mu ^{\pm }e^{\mp }$&0.00&0.72&2&30$\%$
 &6.01&$6.1\times 10^{-4}$&&\\
 $D_{s}^{+}\!\rightarrow \pi ^{-}\mu ^{+}\mu ^{+}$&0.80&0.36&0&27$\%$&1.60
 &$8.2\times 10^{-5}$&$4.3\times 10^{-4}$&\cite{E653}\\
 $D_{s}^{+}\!\rightarrow \pi ^{-}e^{+}e^{+}$&0.00&0.42&1&29$\%$&4.44
 &$6.9\times 10^{-4}$&&\\
 $D_{s}^{+}\!\rightarrow \pi ^{-}\mu ^{+}e^{+}$&0.00&0.36&3&28$\%$&8.21
 &$7.3\times 10^{-4}$&&\\
\hline
\vspace*{-11pt} &     &   &     &     &    &     &    & \\
 $D^{0}\!\rightarrow \mu ^{+}\mu ^{-}$&1.83&0.63&2&6$\%$&3.51
 &$5.2\times 10^{-6}$&$4.1\times 10^{-6}$&\cite{BEATRICE}\\
 $D^{0}\!\rightarrow e^{+}e^{-}$&1.75&0.29&0&9$\%$&1.26
 &$6.2\times 10^{-6}$&$8.2\times 10^{-6}$&\cite{E789}\\
 $D^{0}\!\rightarrow \mu ^{\pm }e^{\mp }$&2.63&0.25&2&7$\%$&3.09
 &$8.1\times 10^{-6}$&$1.7\times 10^{-5}$&\cite{E789}\\
\hline
\end{tabular}
\end{table}

The sources of systematic errors in this analysis included: 
statistical errors from the fit to the normalization sample 
$N_{\mathrm{Norm}}$; statistical errors on the numbers of Monte Carlo 
generated events for both $N_{\mathrm{Norm}}^{\mathrm{MC}}$ and 
$N_{X}^{\mathrm{MC}}$; uncertainties in the calculation of 
misidentification background; and uncertainties in the relative 
efficiency for each mode, including lepton and kaon tagging 
efficiencies. These tagging efficiency uncertainties included: 1) the 
muon counter efficiencies from both Monte Carlo simulation and 
hardware performance; 2) kaon \cerenkov{} identification 
efficiency due to differences in kinematics and modeling between 
data and Monte Carlo simulated events; and 3) the fraction of 
signal events (based on simulations) that would remain outside the 
signal window due to bremsstrahlung tails. 
The larger systematic errors for the $D_{s}^{+}$ modes, 
compared to the $D^{+}$ and $D^{0}$ modes, are due to the uncertainty 
in the branching fraction for the $D_{s}^{+}$ normalization mode.
The sums, taken in quadrature, of these systematic errors are listed 
in Table 1.

\section{Summary}
\noindent
As demonstrated elsewhere,\cite{SCHWARTZ93} one can estimate what new 
mass regions these results can probe. If one, for simplicity, assumes 
that $g_{X^0,X'}=g_W$ then for FCNC, using:
\begin{equation}
m_{X^0}\sim m_W\cdot {\left[ \frac{BR(D^+\to {\bar K}^0 \mu ^+
\nu _\mu)} {BR(D^+\to \pi ^+\mu ^+ \mu ^-)}\cdot (2.2) \right]}^{1/4},
\end{equation}
one estimates that $m_{X^{0}}>800$ GeV/$c^{2}$. Applying a similar 
technique for LFV, and now using the formula:
\begin{equation}
m_{X'}\sim m_W\cdot {\left[ \frac{BR(D^+\to {\bar K}^0 \mu ^+
\nu _\mu )} {BR(D^+\to \pi ^+\mu ^{\pm }e^{\mp })}
\cdot (2.2) \right]}^{1/4},
\end{equation}
one estimates that $m_{X'}>650$ GeV/$c^{2}$, consistent with 
theoretical predictions.\cite{Leurer}

In summary, a ``blind'' analysis of data from Fermilab experiment E791 
was used to obtain upper limits on the dilepton branching 
fractions for flavor-changing neutral current, lepton-number violating, 
and lepton-family violating decays of $D^+$, $D_{s}^{+}$, and $D^0$ 
mesons. No evidence for any of these decays was found. Therefore, upper 
limits on the branching fractions at the 90$\%$ confidence level were 
presented. These limits represent significant improvements over 
previously published results. Eight new $D_{s}^{+}$ search modes were 
reported. 

For the future, FOCUS (Fermilab experiment E831)\cite{Sheldon} will 
shortly be publishing their results, which should improve on the E791 
results. Also, because of interest expressed\cite{Singer} in resonant 
decays of $D^0$s, Fermilab experiment E791 will shortly be publishing 
results of a search for $D^{0}\rightarrow V\ell ^{\pm }\ell ^{\mp }$ 
decays.

\nonumsection{Acknowledgments}
\noindent
I gratefully acknowledge the assistance of Alan Schwartz for the 
calculations of the mass regions probed by these rare and forbidden 
decays. I would also like to express gratitude to Zoltan Ligeti, for 
inviting me to give the talk at the Fermilab Joint Experimental 
Theoretical Seminar that was the basis for this review. This research 
supported under DOE grant DE-FG05-91ER40622.
\vfill
\eject

\nonumsection{References}

\end{document}